\documentclass[10pt,letter,twoside]{aa}

\usepackage{graphicx}
\usepackage{latexsym}
\usepackage{amsmath}
\usepackage{amssymb}
\usepackage{amstext}

\usepackage{natbib}
\bibpunct{(}{)}{;}{a}{}{,}

\begin{document}

    \title{Morphology of synchrotron emission in young supernova remnants}

    \titlerunning{Morphology of synchrotron emission in young SNRs}

    \author{G.~Cassam-Chena{\"\i}\inst{1}$^{\&}$\inst{2}
            \and A.~Decourchelle\inst{2}
            \and J.~Ballet\inst{2}
            \and D. C.~Ellison\inst{1}}

    \authorrunning{G.~Cassam-Chena{\"\i} et al.}

    \institute{Department of Physics, North Carolina State University, Box 8202, Raleigh, NC 27695,
    USA\\\email{gcassam@ncsu.edu}
        \and Service d'Astrophysique, CEA Saclay, 91191 Gif-sur-Yvette, France}

\date{}

\abstract{ In the framework of test-particle and cosmic-ray
modified hydrodynamics, we calculate synchrotron emission radial
profiles in young ejecta-dominated supernova remnants (SNRs)
evolving in an ambient medium which is uniform in density and
magnetic field. We find that, even without any magnetic field
amplification by Raleigh-Taylor instabilities, the radio
synchrotron emission peaks at the contact discontinuity because
the magnetic field is compressed and is larger there than at the
forward shock. The X-ray synchrotron emission sharply drops behind
the forward shock as the highest energy electrons suffer severe
radiative losses.

 }

\maketitle


\section{Introduction\label{sect-intro}}

Shocks in supernova remnants (SNRs) are believed to produce the
majority of the Galactic cosmic-rays (CRs) at least up to the
``knee'' ($\sim 3 \times 10^{15}$ eV). The particle acceleration
mechanism most likely responsible for this is known as diffusive
shock acceleration (DSA) \citep[e.g.,][]{dr83, ble87}. This
mechanism may transfer a large fraction of the ram kinetic energy
(up to $50\%$) into relativistic particles and remove it from the
thermal plasma \citep[see, for example,][]{joe91}.

Convincing observational support for the acceleration of particles
in shell-type SNRs comes from their nonthermal radio and X-ray
emissions due to synchrotron radiation from relativistic GeV and
at least TeV electrons, respectively. In radio and X-rays,
synchrotron-dominated SNRs display various morphologies: for
instance, the synchrotron emission dominates in two bright limbs
in SN 1006 \citep[e.g.,][]{rob04} whereas it is distorted and
complex in RX J1713.7--3946 \citep[e.g.,][]{cad04b}. The detection
and imaging with the \textit{HESS} telescopes of TeV $\gamma$-rays
in RX J1713.7--3946 provides unambiguous evidence for particle
acceleration to very high energies. The $\gamma$-ray morphology in
this remnant is similar to that seen in X-rays \citep{aha04}.

Recent works based on \textit{Chandra} \citep[][ for Cas
A]{vil03a} and \textit{XMM-Newton} \citep[][ for Kepler's
SNR]{cad04a} observations have demonstrated that X-ray synchrotron
emission is also present in ejecta-dominated SNRs and largely
contributes to the continuum emission at the forward shock. This
X-ray emission arises from sharp filaments encircling the SNR's
outer boundary. The observed width of these filaments is a few
arcseconds, and has been used to constrain the magnetic field
intensity just behind the shock\footnote{Magnetic field values are
found to be at least 30 times higher than the typical Galactic
field of $3 \: {\mu}$G and imply that the field has been
amplified, perhaps by the particle acceleration process
\citep{bel01}.} \citep{vil03a, bek03, bev04a, vob05, ba05}.

A number of recent hydrodynamical models, including particle
acceleration and photon emission, have been presented to explain
various features of these observations. \citet{re98} has described
the morphology and spectrum of the synchrotron X-ray emission from
SNRs in the Sedov evolutionary phase. Similar work based on
numerical simulations was done by \citet{vaa04} who take into
account the diffusion of particles. CRs are treated as
test-particles in these studies.

Here, we expand on the work of \citet{re98} by considering young
(ejecta-dominated) SNRs. We investigate the synchrotron emission
morphology, both in radio and X-rays, as well as how it can be
modified by efficient particle acceleration. Our results show that
the radio and X-ray profiles are very different due to the effects
of the magnetic field evolution and synchrotron losses in the
interaction region between the contact discontinuity and the
forward shock. For typical parameters, the radio emission peaks at
the contact discontinuity while the X-ray emission forms
sheet-like structures at the forward shock.

\section{Hydrodynamics and particle acceleration\label{sect-hydro}}

The hydrodynamic evolution of young supernova remnants, including
the backreaction from accelerated particles, can be described by
self-similar solutions if the initial density profiles in the
ejected material (ejecta) and in the ambient medium have power-law
distributions \citep{ch82, ch83}, and if the acceleration
efficiency (\emph{i.e.} the fraction of total ram kinetic energy
going into suprathermal particles) is independent of time.

Here, we use the self-similar model of \citet{ch83} which
considers a thermal gas ($\gamma=5/3$) and the cosmic-ray fluid
($\gamma=4/3$), with the boundary conditions calculated from the
non-linear diffusive shock acceleration (DSA) model of
\citet{bee99} and \citet{elb00} as described in \citet{dee00}.
This acceleration model is an approximate, semi-analytical model
that determines the shock modification and particle spectrum from
thermal to relativistic energies in the plane-wave, steady state
approximation as a function of an arbitrary injection parameter,
$\eta_{\mathrm{inj}}$ (\emph{i.e.} the fraction of total particles
which end up with suprathermal energies). The validity of the
self-similar solutions has been discussed by \citet{dee00} and
direct comparisons between this self-similar model and the more
general CR-hydro model of \citet{eld04} showed good correspondence
for a range of input conditions.

The hydrodynamic evolution provides the shock characteristics
necessary to calculate the particle spectrum at the forward
shock\footnote{We do not consider CR production at the reverse
shock since the magnetic field at the reverse shock may be
considerably smaller than that at the forward shock due to the
dilution by expansion and flux freezing of the progenitor magnetic
field \citep[see][]{eld05}.}, at any time. Once a particle
spectrum has been produced at the shock, it will evolve downstream
because of radiative and adiabatic expansion losses. We assume
that the accelerated particles remain confined to the fluid
element in which they were produced, so adiabatic losses are
determined directly from the fluid element expansion. The basic
power law spectrum produced by DSA, before losses are taken into
account, is modified at the highest energies with a exponential
cutoff, $\exp(-p/p_{\mathrm{max}})$, where $p_{\mathrm{max}}$ is
determined by matching either the acceleration time to the shock
age or the upstream diffusive length to some fraction of the shock
radius. In our simulation, the electron-to-proton density ratio at
relativistic energies, $(e/p)_{\mathrm{rel}}$, is set equal to
$0.01$ \citep[see][]{elb00}.

Unless explicitly stated, our numerical examples are given for the
following supernova parameters: $M_{\mathrm{ej}} = 5 \;
\mathrm{M}_{\odot}$ for the ejected mass, $E_{51} = 1$ where
$E_{51}$ is the kinetic energy of the ejecta in units of $10^{51}$
ergs and $n=9$, where $n$ is the index of the initial power-law
density profile in the ejecta ($\rho \propto r^{-n}$). In our
simulations, the SNR age is $t_{f} = 400$ years and the shock
velocity at the forward shock is $v_{s} \simeq 5 \times 10^{3} \:
\mathrm{km/s}$. For the ambient medium parameters, we take a
magnetic field $B_0 = 10 \: \mu\mathrm{G}$, a density $n_0 = 0.1
\; \mathrm{cm}^{-3}$, an ambient gas pressure $p_{\mathrm{g},0}/k
= 2 \: 300 \; \mathrm{K} \: \mathrm{cm}^{-3}$ and $s=0$, where $s$
is the index of its initial power-law density profile ($\rho
\propto r^{-s}$). The case $s=0$ corresponds to a uniform
interstellar medium ($s=2$ describes a stellar wind).

In the next section,  we discuss the importance of the magnetic
field for the synchrotron emission and particle acceleration. We
do not, however, explicitly include the dynamical influence of the
magnetic field on the hydrodynamics.

\section{Results\label{sect-results}}

    \subsection{Magnetic field\label{subsect-MF}}

To track the synchrotron losses, we are interested in the temporal
evolution of the magnetic field behind the shock. We assume the
magnetic field to be simply compressed at the shock and passively
carried by the flow, frozen in the plasma, so that it evolves
conserving flux. In this simple 1-D approach, we do not consider
any production of the SNR magnetic field, for instance, by
hydrodynamical instabilities which is an additional effect. As for
the magnetic field ahead of the forward shock, it is assumed to be
isotropic and fully turbulent. Appendix \ref{app-MFevol} (see the
on-line version) shows how to compute the magnetic field profile
for self-similar solutions in both test-particle and nonlinear
particle acceleration cases.

        \subsubsection{Test-Particle limit\label{Test-particle}}

We first discuss the behavior of the normal and tangential
components of the magnetic field in the test-particle case where
the backreaction of the accelerated particles is neglected.

\begin{figure}[t]
\centering
\includegraphics[width=9cm]{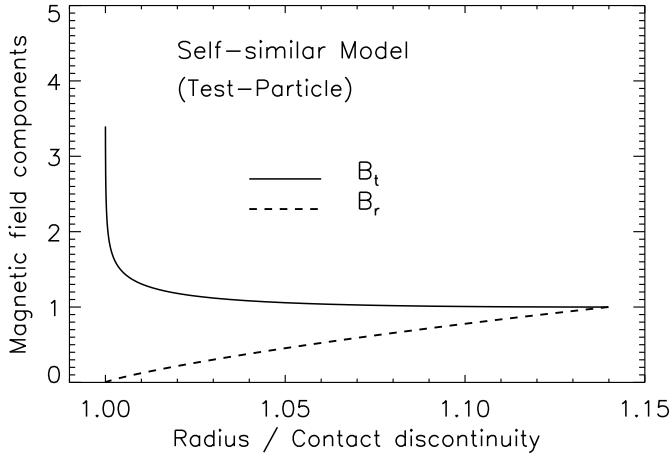}
\caption{Radial profile of the normal ($B_r$) and tangential
($B_t$) components of the magnetic field in a test-particle
self-similar model. Each component is normalized to the forward
shock.} \label{fig-B-profile-TP-n9s0q0}
\end{figure}

When the SNR evolves in an ambient medium which is uniform in
density and magnetic field, the expansion and flux freezing
generally cause the tangential component of the magnetic field to
increase at the contact discontinuity whereas the normal component
falls to zero (Fig. \ref{fig-B-profile-TP-n9s0q0}). As a result,
the magnetic field profile is dominated by the tangential
component.

One has often invoked hydrodynamic instabilities to explain the
magnetic field increase at the interface between the shocked
ejecta and the shocked ambient medium \citep{jun95}. The numerical
simulations of \citet{jun96} have shown that the magnetic field
could be amplified by a factor 60 by Rayleigh-Taylor and
Kelvin-Helmholtz instabilities. Here, we note that simple
advection of the magnetic field already predicts amplification by
a factor 5 (Table \ref{Tab-sigmaB} top, $n=9$).

We note that, if the SNR evolves in a wind with a decreasing
initial density profile, advection goes the other way (diluting
the magnetic field instead of amplifying it). But when both the
ambient density and magnetic field decrease with radius, as would
be the case for a pre-supernova stellar wind, the magnetic field
is larger close to the contact discontinuity than at the forward
shock (by a factor of $\sim 1000$ in some cases). This is because
the dilution of the advected magnetic field is negligible compared
to the fact that the ambient magnetic field was much larger at
early times.

        \subsubsection{Nonlinear Particle Acceleration}

\begin{figure}[t]
\centering
\includegraphics[width=9cm]{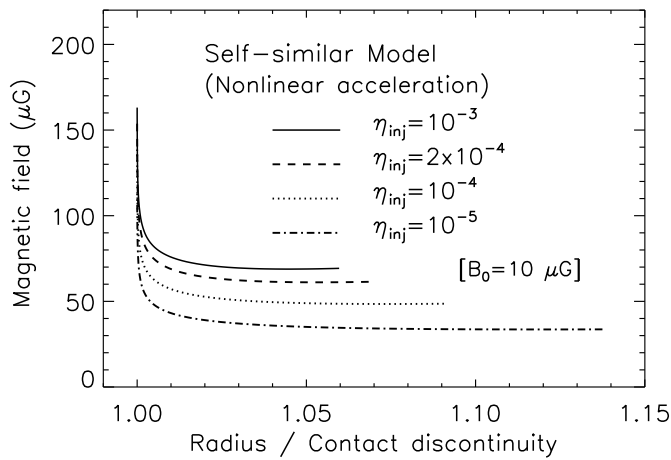} \caption{Magnetic field radial
profile for different values of the injection efficiency,
$\eta_{\mathrm{inj}}$, when the hydrodynamics is coupled with the
non-linear DSA model. The width of the shocked region is smaller
and smaller as the feedback of the accelerated particles on the
SNR dynamics increases. } \label{fig-B-profile-NL}
\end{figure}

We now consider the behavior of the normal and tangential
components of the magnetic field in the nonlinear case where the
backreaction of the accelerated particles on the shock is taken
into account.

In the ideal non-linear case, where the acceleration is
instantaneous, the magnetic field diverges at the contact
discontinuity because of its tangential component, whatever the
injection efficiency is, as in the test-particle case. However,
the contrast between the magnetic field in a given fluid element
and the one just behind the shock, will be always smaller than in
the test-particle case (see Table \ref{Tab-sigmaB}). Figure
\ref{fig-B-profile-NL} shows the profile of the total downstream
magnetic field for different values of the injection efficiency.
Table \ref{Tab-rtot_etainj} shows the associated compression ratio
and immediate post-shock magnetic field.

\begin{table}[t]
\centering
\begin{tabular}{lcccc}
\hline \hline $\eta_{\mathrm{inj}}$ &  $10^{-3}$ & $2 \times
10^{-4}$ & $10^{-4}$ & $10^{-5}$
\\ \hline
$r_{\mathrm{tot}}$  & 8.5 & 7.5 & 5.9 & 4.1 \\
$B_{s}$ (${\mu}\mathrm{G}$) & 69 & 61 & 49 & 34 \\ \hline
\end{tabular}
\caption{Compression ratio, $r_{\mathrm{tot}}$, and downstream
magnetic field, $B_{s}$, at the forward shock obtained for
different injection, $\eta_{\mathrm{inj}}$. The magnetic field
compression ratio is given by $r_{\mathrm{B}} \equiv B_{s}/B_{0} =
\sqrt{1/3+2 \: r_{\mathrm{tot}}^{2}/3}$ and $B_{0}= 10 \:
{\mu}\mathrm{G}$ here.} \label{Tab-rtot_etainj}
\end{table}

    \subsection{Synchrotron emission\label{subsect-syn-emis}}

\begin{figure}[t]
\centering
\includegraphics[bb= 0 60 504 360,clip,width=9cm]{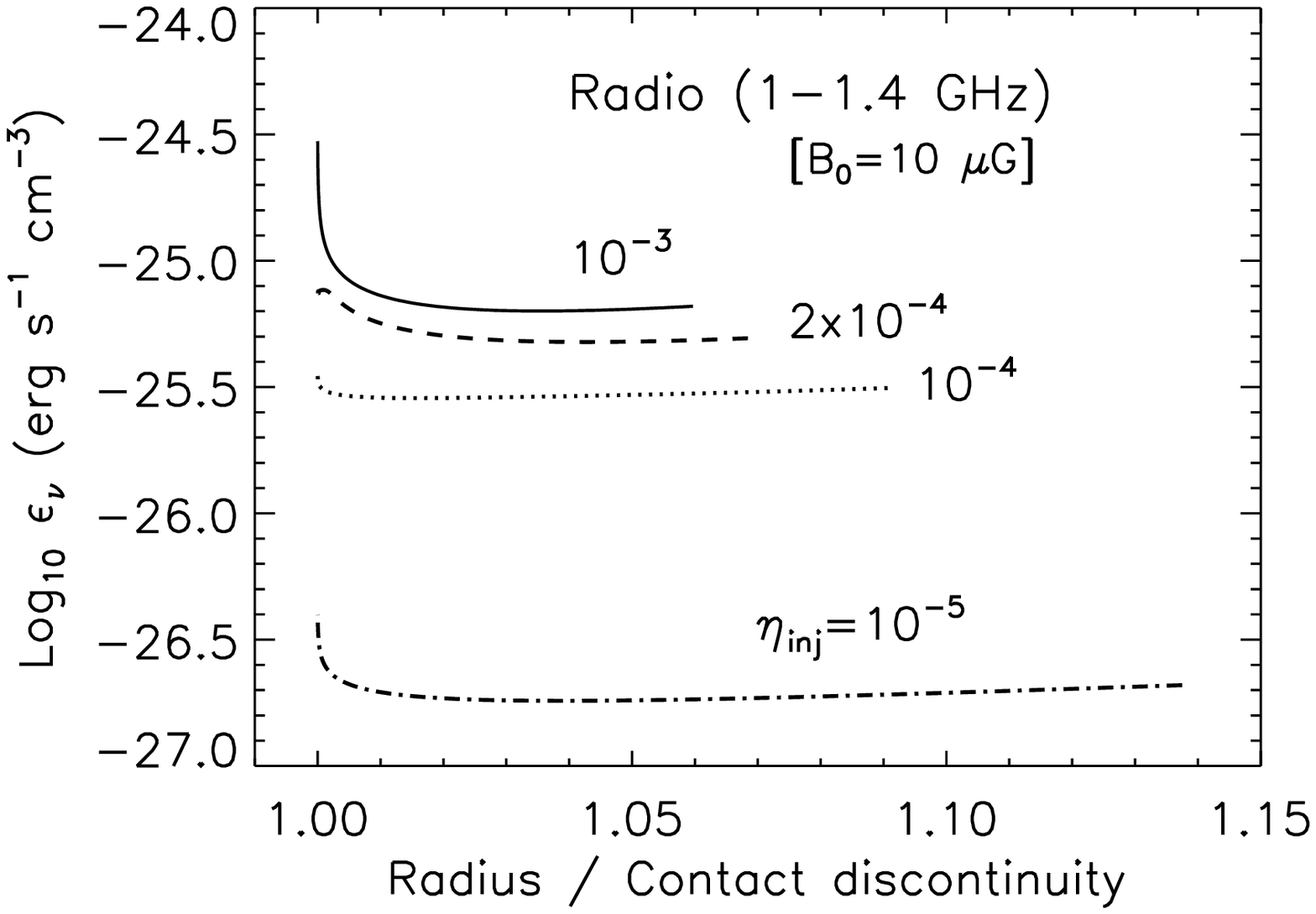}
\includegraphics[width=9cm]{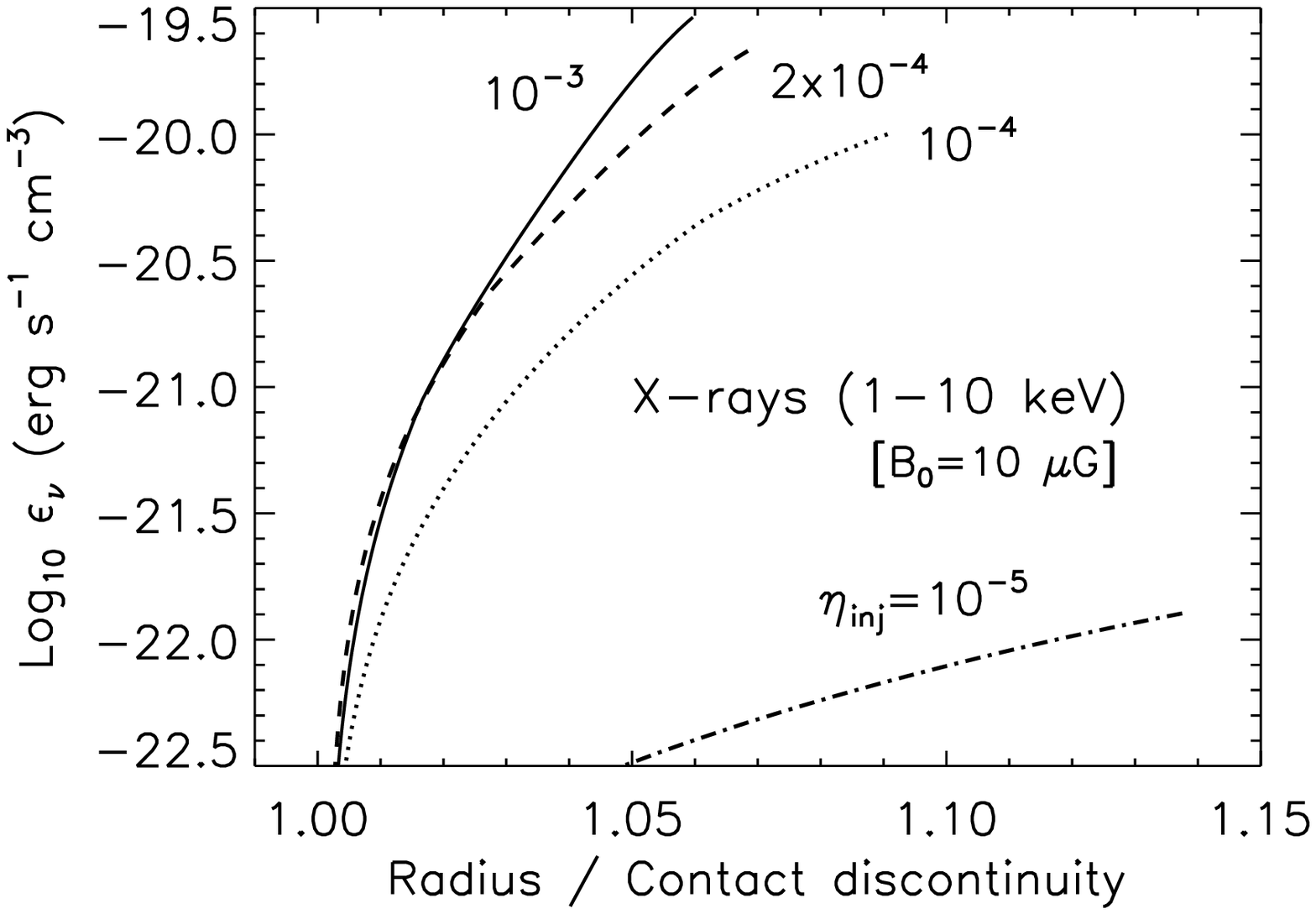}
\caption{Radio (\textit{top panel}) and X-ray (\textit{bottom
panel}) synchrotron volume emissivity, $\epsilon_{\nu}$, radial
profile for different injection efficiencies. }
\label{fig-synch-etainj}
\end{figure}

\begin{figure}[t]
\centering
\includegraphics[bb= 0 60 504 360,clip,width=9cm]{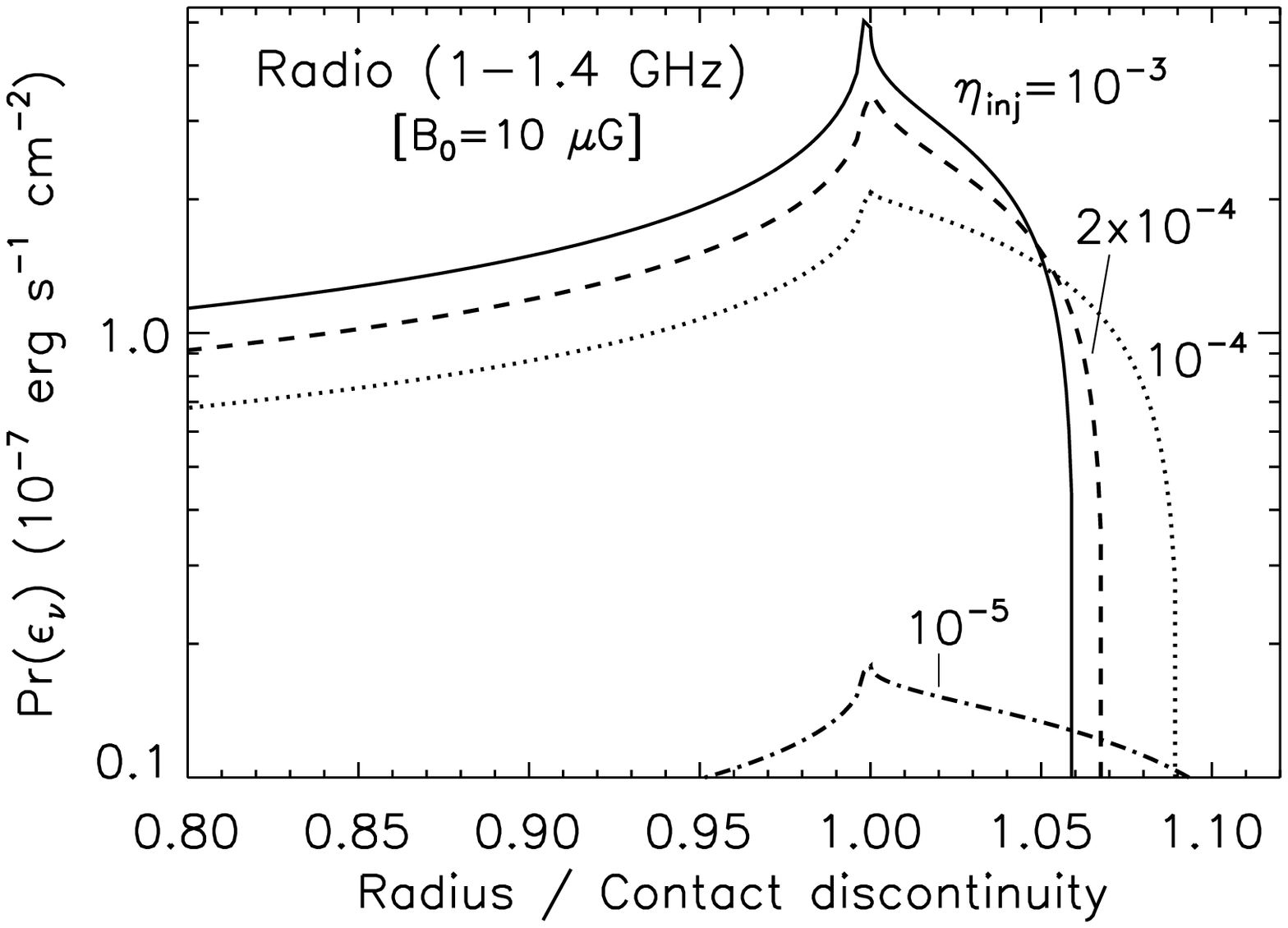}
\includegraphics[bb= 0 0 504 360,clip,width=9cm]{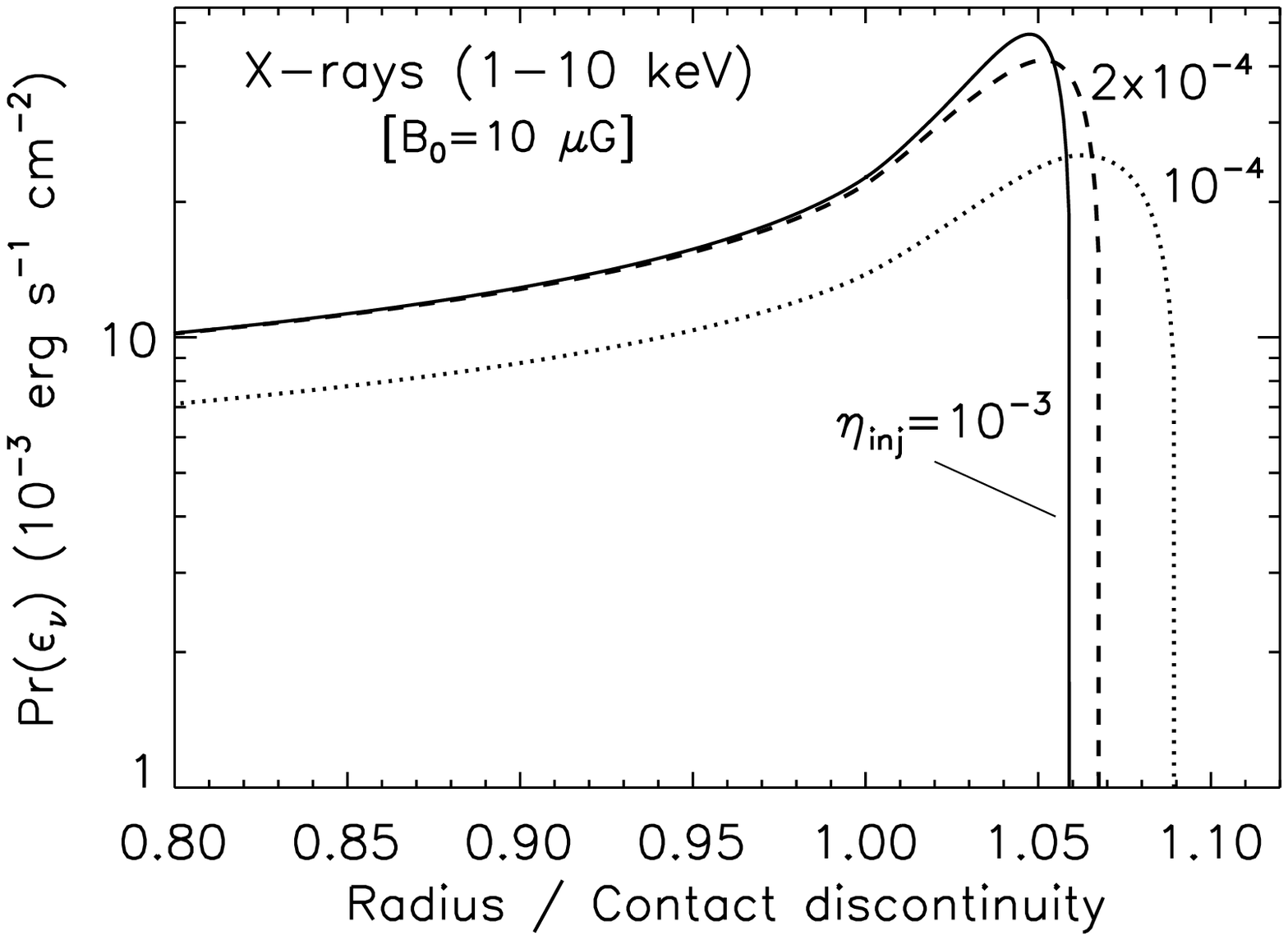}
\caption{Radio (\textit{top panel}) and X-ray (\textit{bottom
panel}) synchrotron volume emissivity radial profile after
projection onto the line-of-sight for different injection
efficiencies. Emission from particles accelerated at the reverse
shock is assumed to be negligible.} \label{fig-synchproj-etainj}
\end{figure}

Once the magnetic field structure and the particle spectrum
(attached to a fluid element) modified by the radiative and
adiabatic expansion losses as computed in \citet{re98} are known,
we compute the synchrotron emission \citep{ryl79}, averaged over
the pitch-angle, in any energy band\footnote{We did not calculate
the synchrotron emission from the precursor.}.

Figure \ref{fig-synch-etainj} shows the radial profiles of the
synchrotron emission in the radio (top panel) and X-ray (bottom
panel) domains for different injection efficiencies,
$\eta_{\mathrm{inj}}$. An increase in the injection efficiency not
only provides a larger number of accelerated electrons, but also a
larger compression of the downstream magnetic field (see Table
\ref{Tab-rtot_etainj}) and a narrower interaction region. These
effects combine to produce enhanced synchrotron emission as the
injection increases.

The radio synchrotron emission is produced by GeV electrons which
are not affected by radiative losses. Consequently, the radio
synchrotron emission critically depends on the final magnetic
field profile (Fig. \ref{fig-B-profile-NL}) and, therefore, peaks
at the contact discontinuity. In contrast, the X-ray synchrotron
emission is produced by the highest momentum electrons ($\sim
10^{3-5} \: m_{\mathrm{p}} \: c$) which, depending on the
downstream field strength,  may suffer radiative losses. The high
energy electrons that have been accelerated at the earliest time
have suffered strong synchrotron losses as they were advected
behind the shock. Because of this, they are not numerous enough at
the end to radiate in the X-ray regime despite a strong magnetic
field. As a result, the X-ray synchrotron emission rapidly
decreases behind the shock. The X-ray profile becomes sharper when
the injection efficiency increases because it provides larger
compression of the downstream magnetic field and then stronger
synchrotron losses.

Figure \ref{fig-synchproj-etainj} shows the synchrotron emission
after integration along the line-of-sight. The radial profile of
the radio emission (top panel) shows a peak at the contact
discontinuity. The radial profile of the X-ray projected
synchrotron emission (bottom panel) shows bright rims just behind
the forward shock whose width decreases as the injection
efficiency increases.

\section{Discussion and Conclusion\label{sect-concl}}

We have computed the radio and X-ray synchrotron emission in young
ejecta-dominated SNRs. This has been done using a one dimensional,
self-similar hydrodynamical calculation coupled with a non-linear
diffusive shock acceleration model, and taking into account the
adiabatic and radiative losses of the electron spectrum during its
advection in the remnant.

We show that the morphology of the synchrotron emission in young
ejecta-dominated SNRs is very different in radio and X-ray. This
is the result of the increased magnetic field toward the contact
discontinuity, to which only low energy electrons that emit radio
are sensitive, while the high energy electrons emitting X-rays
experience strong radiative losses and are mostly dependent on the
post-shock magnetic field.

Briefly, the radio synchrotron emission increases as one moves
from the forward shock toward the contact discontinuity due to a
compression of the magnetic field (particularly its tangential
component), assuming both uniform ambient density and upstream
magnetic field. Such a compression naturally results from the
dynamical evolution of the SNR. In contrast, because of the
radiative losses, the X-ray synchrotron emission decreases behind
the forward shock and forms sheet-like structures after
line-of-sight projection. Their widths decrease as the
acceleration becomes more efficient.

The morphology of the radio synchrotron emission obtained for the
young ejecta-dominated stage of SNRs will differ from that of SNRs
in the Sedov phase (but not in X-ray). Indeed, \citet{re98} has
shown that both the normal and tangential components of the
magnetic field decrease behind the forward shock in the Sedov
phase and, as a result, we expect the radio synchrotron emission
to decrease behind the shock (however, less rapidly than the X-ray
synchrotron emission since the radio electrons do not experience
radiative losses).

Our model qualitatively reproduces the main features of the radio
and X-ray observations of emission in young ejecta-dominated SNRs
(e.g., Tycho and Kepler), \emph{i.e.} bright radio synchrotron
emission at the interface between the shocked ejecta and ambient
medium, and a narrow filament of X-ray emission at the forward
shock. However, this model is unable to reproduce the thin radio
filaments observed at the forward shock in some SNRs \citep[for
instance those seen in Tycho's SNR,][]{div91}.

We note that extensions of this work to cases with exponential
ejecta profiles and/or SNRs evolving in a pre-supernova stellar
wind with varying magnetic fields, cannot be done with
self-similar solutions. These cases can be calculated in the
numerical CR-modified hydrodynamical model described in
\citet{eld05} and this work is in progress \citep{elc05}.



\bibliographystyle{aa} 
\bibliography{biblio_article3} 

\Online

\appendix

\section{Magnetic field evolution\label{app-MFevol}}

The evolution of the normal (subscript $r$) and tangential
(subscript $t$) components of the magnetic field at the downstream
position, $B$, is given by \citep{rec81}:
\begin{eqnarray}\label{Br(r)}
B_{r}(r) & = & B_{r,j} \: \left( \frac{r}{r_j} \right)^{-2} \\
\label{Bt(r)} B_{t}(r) & = & B_{t,j} \: \frac{\rho}{\rho_j} \:
\frac{r}{r_j},
\end{eqnarray}
and the total magnetic field is simply \citep{re98}:
\begin{equation}
    B(r) = \left( B_{r}(r)^2 + B_t(r)^2 \right)^{1/2}.
\end{equation}
In these equations, $r$ and $\rho$ are, respectively, the radius
and density of a fluid element at the current time that was
shocked at the previous time $t_j$. At time $t_j$, the fluid
element was just behind the shock at the radius $r_{j}$, with a
density $\rho_j$ and a magnetic field $B_{j}$.

We assume that the upstream magnetic field at time $t_{j}$,
$B_{0,j}$ is isotropic and fully turbulent so that the components
of the immediate post-shock magnetic field $B_{j}$ in Eqs
(\ref{Br(r)}) and (\ref{Bt(r)}) are given on average by
\citep{bek02}:
\begin{eqnarray}\label{Brj-Btj}\label{Brj}
B_{r,j} &=& 1/ \sqrt{3} \: B_{0,j}\\ \label{Btj} B_{t,j} &=&
\sqrt{2/3} \: r_{\mathrm{tot}} \: B_{0,j}.
\end{eqnarray}
where $r_{\mathrm{tot}}$ is the shock compression ratio. In the
self-similar approach, $r_{\mathrm{tot}}$ is assumed independent
of time \cite[see][ for details]{dee00}.

We consider that the current magnetic field upstream of the
forward shock, $B_{0,s}$, can behave like:
\begin{equation}\label{B-wind}
    B_{0,s} = B_{0,j} \left( \frac{r_{s}}{r_{j}} \right)^{-q}
\end{equation}
where $r_{s}$ is the current shock radius. If the magnetic field
is uniform, the index $q$ is equal to 0. In a stellar wind
($s=2$), the magnetic field profile may be decreasing yielding
$q=1$ \citep{lyp04} or $q=2$ if we assume that it is frozen in the
plasma.

We define the magnetic field contrast factor, $\sigma_{B} \equiv
B/B_{s}$, as the ratio between the current magnetic field in a
fluid element, $B$, and the current one just behind the shock,
$B_{s}$. We have:
\begin{equation}\label{sigmaB}
    \sigma_{B}  = \left(
    \frac{ \sigma_{B_{r}}^{2} + 2 \: r_{\mathrm{tot}}^{2} \:
    \sigma_{B_{t}}^{2}} { 1 + 2 \: r_{\mathrm{tot}}^{2}}
    \right)^{1/2}
\end{equation}
where $\sigma_{B_{r}} \equiv B_{r}/B_{r,s}$ and $\sigma_{B_{t}}
\equiv B_{t}/B_{t,s}$ are the magnetic field contrast factors of
the normal and tangential components of the field, respectively.
The components $B_{r,s}$ and $B_{t,s}$ obey the same relation as
in Eqs (\ref{Brj}) and (\ref{Btj}).

\subsection{Test-Particle limit\label{Test-particle}}

Assuming adiabaticity of the thermal gas, the magnetic field
contrast factors of the normal and tangential components of the
field are given by:
\begin{eqnarray} \label{sigmaBr}
  \sigma_{B_{r}}  &=& \left( \frac{R_s}{R} \right)^2 \: \left( \frac{v_{j}}{v_{s}}
  \right)^{\beta_{r}}\\\label{sigmaBt}
  \sigma_{B_{t}} &=& \left( \frac{P_{\mathrm{g},s}}{P_{\mathrm{g}}} \right)^{-3/5}
\: \left( \frac{R_s}{R} \right)^{-(11-3s)/5}
  \:\left( \frac{v_{j}}{v_{s}} \right)^{\beta_{t}}
\end{eqnarray}
where the indexes $\beta_{r}$ and $\beta_{t}$ are given by:
\begin{eqnarray}\label{betar}
  \beta_{r} &=& \left( q-2 \right) \: \frac{n-3}{3-s} \\ \label{betat}
  \beta_{t} &=& \frac{5 n - 33 - 3 s (n-5)}{5(3-s)} + q \: \frac{n-3}{3-s}.
\end{eqnarray}
In Eqs (\ref{sigmaBr}) and (\ref{sigmaBt}), $R_s/R$ and
$P_{\mathrm{g},s}/P_{\mathrm{g}}$ are the ratio of the
self-similar radii and thermal gas pressures, respectively,
between the shock (subscript $s$) and a fluid element
\citep[see][]{ch82}. They depend on $n$ and $s$, but also weakly
on $v_{j}/v_{s}$ where $v_{s}$ and $v_{j}$ are the current shock
velocity and the shock velocity at the time $t_j$, respectively.

In the framework of these self-similar solutions, the forward
shock velocity tends to infinity at early times, corresponding to
fluid elements close to the contact discontinuity at the current
time. To limit the maximum velocity to a realistic value, we look
at the value of $\sigma_{B}$ for a shock velocity ratio
$v_{j}/v_{s} = 10$. For the typical forward shock velocity $v_{s}$
that we have used for the numerical application, the initial
velocity corresponds to $v_{j} \simeq 5 \times 10^{4} \:
\mathrm{km/s}$. This shock velocity is the criterion used to
define the radial position of the oldest fluid element that is
currently located close to the contact discontinuity.

Here, we consider the case of both an uniform ambient medium
($s=0$) and upstream magnetic field ($q=0$). Under this
assumption, $B_{s}=B_{j}$, since $r_{\mathrm{tot}}$ is constant
with time. Then, the magnetic field contrast factor, $\sigma_{B}$
is equal to $B/B_{j}$ and can be viewed as a compression or a
dilution factor. Table \ref{Tab-sigmaB} (top) gives the contrast
$\sigma_{B}$ for different values of $n$.

\begin{table*}[t]
\centering
\begin{tabular}{lccccccc}
\hline \hline  & \multicolumn{1}{l}{$n$} &
\multicolumn{1}{c}{$R_{s}/R$} &
\multicolumn{2}{c}{$P_{\mathrm{g},s}/P_{\mathrm{g}}$} &
\multicolumn{2}{c}{$\sigma_{B}$} & \multicolumn{1}{c}{$v_{s}$
(km/s)} \\ \hline
& 7 & 1.181 & \multicolumn{2}{c}{0.964} & \multicolumn{2}{c}{0.95} & 4840  \\
\multicolumn{1}{l}{Test-Particle} & 9 & 1.140 & \multicolumn{2}{c}{0.885} & \multicolumn{2}{c}{5.0} & 4850 \\
& 12 & 1.121 & \multicolumn{2}{c}{0.836} & \multicolumn{2}{c}{54}
& 4940 \\
 \hline  &
\multicolumn{1}{l}{$n$} & \multicolumn{1}{c}{$R_{s}/R$} &
\multicolumn{1}{c}{$P_{s}/P$} &
\multicolumn{1}{c}{$P_{\mathrm{c},s}/P$} &
\multicolumn{1}{c}{$\min(\sigma_{B})$} &
\multicolumn{1}{c}{$\max(\sigma_{B})$} &
\multicolumn{1}{c}{$v_{s}$ (km/s)} \\ \hline
& 7 & 1.080 & 1.135 & 0.806 & 0.50 & 0.66 & 4370 \\
\multicolumn{1}{l}{Nonlinear DSA} & 9 & 1.060 & 1.045 & 0.754 & 2.6 & 3.4 & 4470 \\
& 12 & 1.051 & 0.988 & 0.714 & 28 & 36 & 4610 \\
\hline
\end{tabular}
\caption{Test-Particle: Magnetic field contrast factor,
$\sigma_{B}$, for a velocity ratio $v_{j}/v_{s}=10$, radius ratio
and self-similar thermal gas pressure ratio \citep[see][]{ch82}
and shock velocity, for different values of the indexes $n$ ($s=0$
and $q=0$). Nonlinear DSA: same as Test-Particle but with lower
and upper limits on the magnetic field contrast factor,
$\sigma_{B}$, with $\eta_{\mathrm{inj}} = 10^{-3}$ (see Sect.
\ref{sect-Nonlinear-DSA} for more explanations).}
\label{Tab-sigmaB}
\end{table*}

\subsection{Nonlinear Particle Acceleration\label{sect-Nonlinear-DSA}}

In the ideal non-linear case, where the acceleration is
instantaneous and efficient, the thermal gas pressure falls to
zero at the contact discontinuity while the relativistic gas
pressure goes to infinity. Hence, the contrast factor of the
tangential field component, $\sigma_{B_{t}}$, given by Eq.
(\ref{sigmaBt}), obtained in the test-particle limit, is not
defined when $v_{j}/v_{s}$ tends to infinity.

However, the contrast of the tangential component of the magnetic
field can also be found by using the adiabaticity of the
relativistic gas:
\begin{eqnarray} \label{sigmaBt-cosmic}
  \sigma_{B_{t}} &=& \left( \frac{P_{\mathrm{c},s}}{P_{\mathrm{c}}} \right)^{-3/4}
\: \left( \frac{R_s}{R} \right)^{-(10-3s)/4}
  \:\left( \frac{v_{j}}{v_{s}} \right)^{\beta_{t}'}
\end{eqnarray}
where the index $\beta_{t}'$ is given by:
\begin{eqnarray}\label{betat-cosmic}
  \beta_{t}' &=& \frac{4 n - 30 - 3 s (n-5)}{4(3-s)} + q \: \frac{n-3}{3-s}.
\end{eqnarray}
In Eq. (\ref{sigmaBt-cosmic}), $P_{\mathrm{c},s}/P_{\mathrm{c}}$
is the ratio of the self-similar relativistic gas pressures
between the shock (subscript $s$) and a fluid element. This ratio
depends on $n$, $s$, and $v_{j}/v_{s}$. The contrast of the normal
field component, $\sigma_{B_{r}}$, is still given by Eq.
(\ref{sigmaBr}). The asymptotic behavior of the contrast factor,
$\sigma_{B_{t}}$, can be derived from Eq. (\ref{sigmaBt-cosmic})
because the relativistic gas pressure does not tend to zero at the
contact discontinuity.

Because the thermal gas pressure vanishes as we approach the
contact discontinuity in the case of ideal particle acceleration,
i.e., when the acceleration is instantaneous and efficient, the
contrast of the tangential field component, $\sigma_{B_{t}}$, will
always be smaller than in the test-particle case where the thermal
gas pressure rapidly tends to a constant (see Eq. \ref{sigmaBt}).
Table \ref{Tab-sigmaB} (bottom) gives the lower and upper limits
on the magnetic field contrast factor, $\sigma_{B}$, in the case
of ideal nonlinear particle acceleration for
$\eta_{\mathrm{inj}}=10^{-3}$ and for different values of $n$ when
both the ambient medium and upstream magnetic field are uniform
($s=0$ and $q=0$). The lower and upper limits on $\sigma_{B}$ are
obtained by replacing in Eq. (\ref{sigmaBt-cosmic}) the ratio of
the self-similar relativistic gas pressures,
$P_{\mathrm{c},s}/P_{\mathrm{c}}$, by the ratio of the
self-similar total gas pressures, $P_{s}/P \equiv
(P_{\mathrm{c},s}+P_{\mathrm{g},s})/(P_{\mathrm{c}}+P_{\mathrm{g}})$,
and by the ratio between the self-similar relativistic gas
pressure at the shock and the self-similar total gas pressure,
$P_{c,s}/P$, respectively.

However, for an injection efficiency lower than $\sim 5 \times
10^{-4}$, the acceleration is not efficient enough for the shock
to be modified at the beginning of the evolution. In that case,
the fluid elements that have been shocked at the earliest times
are still dominated by the thermal gas so that test-particle
solutions could still apply locally.

\end{document}